\documentclass[
reprint,
superscriptaddress,
 amsmath,amssymb,
aps,
]{revtex4-1}
\usepackage{tikz}
\usepackage{braket}
\usepackage{graphicx}
\usepackage{dcolumn}
\usepackage{bm}
\usepackage[mathlines]{lineno}

\begin{document}

\preprint{APS/123-QED}

\title{Fast data-driven spectrometer with direct measurement of time and frequency for multiple single photons}

\author{Jakub Jirsa}
\affiliation{
Faculty of Nuclear Sciences and Physical Engineering, Czech Technical University, 115 19 Prague, Czech Republic
}
\affiliation{
Faculty of Electrical Engineering, Czech Technical University, 166 27 Prague, Czech Republic
}
\author{Sergei Kulkov}
\affiliation{
Faculty of Nuclear Sciences and Physical Engineering, Czech Technical University, 115 19 Prague, Czech Republic
}
\author{Raphael A. Abrahao}
\email{rakelabra@bnl.gov}
\affiliation{
Brookhaven National Laboratory,
Upton NY 11973, USA 
}
\author{Jesse Crawford}
\affiliation{
Brookhaven National Laboratory,
Upton NY 11973, USA 
}
\author{Aaron Mueninghoff}
\affiliation{
Stony Brook University, Stony Brook NY 11794, USA 
}
\author{Ermanno Bernasconi}
\affiliation{
École polytechnique fédérale de Lausanne (EPFL), CH-2002 Neuchâtel, Switzerland
}
\author{Claudio Bruschini}
\affiliation{
École polytechnique fédérale de Lausanne (EPFL), CH-2002 Neuchâtel, Switzerland
}
\author{Samuel Burri}
\affiliation{
École polytechnique fédérale de Lausanne (EPFL), CH-2002 Neuchâtel, Switzerland
}
\author{Stephen Vintskevich}
\affiliation{
Technology Innovation Institute,
Abu Dhabi, United Arab Emirates
}
\author{Michal Marcisovsky}
\affiliation{
Faculty of Nuclear Sciences and Physical Engineering, Czech Technical University, 115 19 Prague, Czech Republic
}
\author{Edoardo Charbon}
\affiliation{
École polytechnique fédérale de Lausanne (EPFL), CH-2002 Neuchâtel, Switzerland
}
\author{Andrei Nomerotski}
\email{nomerot@gmail.com}
\affiliation{
Stony Brook University, Stony Brook NY 11794, USA 
}

\date{\today}

\begin{abstract}
We present a single-photon-sensitive spectrometer, based on a linear array of 512 single-photon avalanche diode detectors, with 0.04 nm spectral and 40 ps temporal resolutions.  We employ a fast data-driven operation that allows direct measurement of time and frequency for simultaneous single photons, time- and frequency-stamping each single-photon detection. Our results combine excellent temporal and spectral resolution. We benchmark our result against the Heisenberg Uncertainty Principle limit of $\hbar/2$ for time and energy, and we are only a factor of 10 above it, despite the simplicity of our experimental setup, including room temperature operation. This work opens numerous applications in both classical and quantum photonics, especially when both spectral and temporal properties of single photons are exploited.
\end{abstract}

\maketitle

\section{\label{sec:Intro}Introduction}

Despite remarkable advances in single-photon detection, modern optical technologies are reaching their limits when high spectral and temporal resolutions are simultaneously needed. This is especially true for applications desiring to approach the limits governed by the Heisenberg Uncertainty Principle (HUP), which can be understood as a consequence of the wave-particle duality in quantum mechanics. We address some of these critical challenges in achieving single-photon detection sensitivity together with excellent spectral (sub-nanometer in wavelength) and temporal (picosecond scale) resolutions. 

Different kinds of fast single-photon detectors have different technical performance measures and focus on different tasks. Common types of fast detectors are single-photon avalanche diodes (SPAD)~\cite{sze2008semiconductor,Optica2020_EPFL,burri2014, Shawkat2018}, superconducting nanowire single-photon detectors (SNSPD)~\cite{Nanowire_NatPHt_2013,NatComm_Yale_spec_nanowire,NatPhot_JPL_spec_nanowire,NatPhot_nanowire_2022_Yale,Loopholefree_Bell_NIST_nanowire,NatPhot_geoff_nanowire_2017}, transition edge sensors (TES)~\cite{Lita_NIST_OE_TES,TES_UQ_arxiv,NatPht_2022_TES,PRL2019_Lewis_TES,LoopholefreeBell_Vienna_TES} and variations of streaking detectors~\cite{Komura_2009, Korobkin1969, Chollet2008,Howorth2016, Kornienko2023}. SPAD detectors are widely used in quantum photonics due to their user-friendly operation, reasonable photon detection efficiency, and cost. Additionally, SPAD sensors operate at room temperature, an important aspect of their ease of use. Production of SPADs in a monolithic complementary metal-oxide-semiconductor (CMOS) process enables better electronics integration and fast digital control interface, while also supporting outstanding temporal resolution and scalability \cite{LinoSPAD_epfl_2016}. Arranging the SPAD detectors in a linear array with easy access to individual pixels allows one to employ the external resources of field programmable gate arrays (FPGA) to perform time stamping and other digital operations. Combining these advances with excellent spectral resolution makes this quantum photonics tool highly compelling.

Such a detector can unlock new uses and applications in both classical and quantum optics, especially when considering spectral binning and wide bandwidth applications. Here we present a spectrometer based on the LinoSPAD2 sensor \cite{milanese2023linospad2,bruschini2023} and test its features, hence confirming its qualities. One of the scientific motivations for the fast spectrometer is to develop new techniques for quantum-assisted optical interferometry for astrophysical applications \cite{Stankus2022,Crawford2023,Brown2023,Brown2022,Nomerotski20}, which benefits from spectral binning. For instance, to observe the Hanbury Brown-Twiss effect with good visibility~\cite{HBT_original,HBT1956_correlation,Stankus2022,Crawford2023}, one needs good temporal and spectral resolutions, as poor resolution in either case will reduce visibility. The spectral binning technique essentially allows one to perform multiple measurements together, because each frequency can be treated as an independent measurement. It also opens the door to other experiments with broadband sources, provided that each frequency is independently measured in the end.
The spectrometer we are reporting could also be used for single-photon source characterization~\cite{SPDC_general,JMP_2018_Joint-spectral-characterization} and optical coherence tomography~\cite{optical_coherence_tomography2004}. Other potential applications of the spectrometer are in spectroscopy, atomic physics, and quantum photonics.

The demonstrated performance extends beyond previous works that only focused on one of the two aspects, either temporal or spectral resolution. Moreover, in the presented spectrometer, time and frequency measurements can be performed for multiple photons, since each pixel is single-photon sensitive.

\section{Results}

\subsection{Fast single-photon-sensitive spectrometer}

\begin{figure}[]
    \centering
    \includegraphics[width=1.0\linewidth]{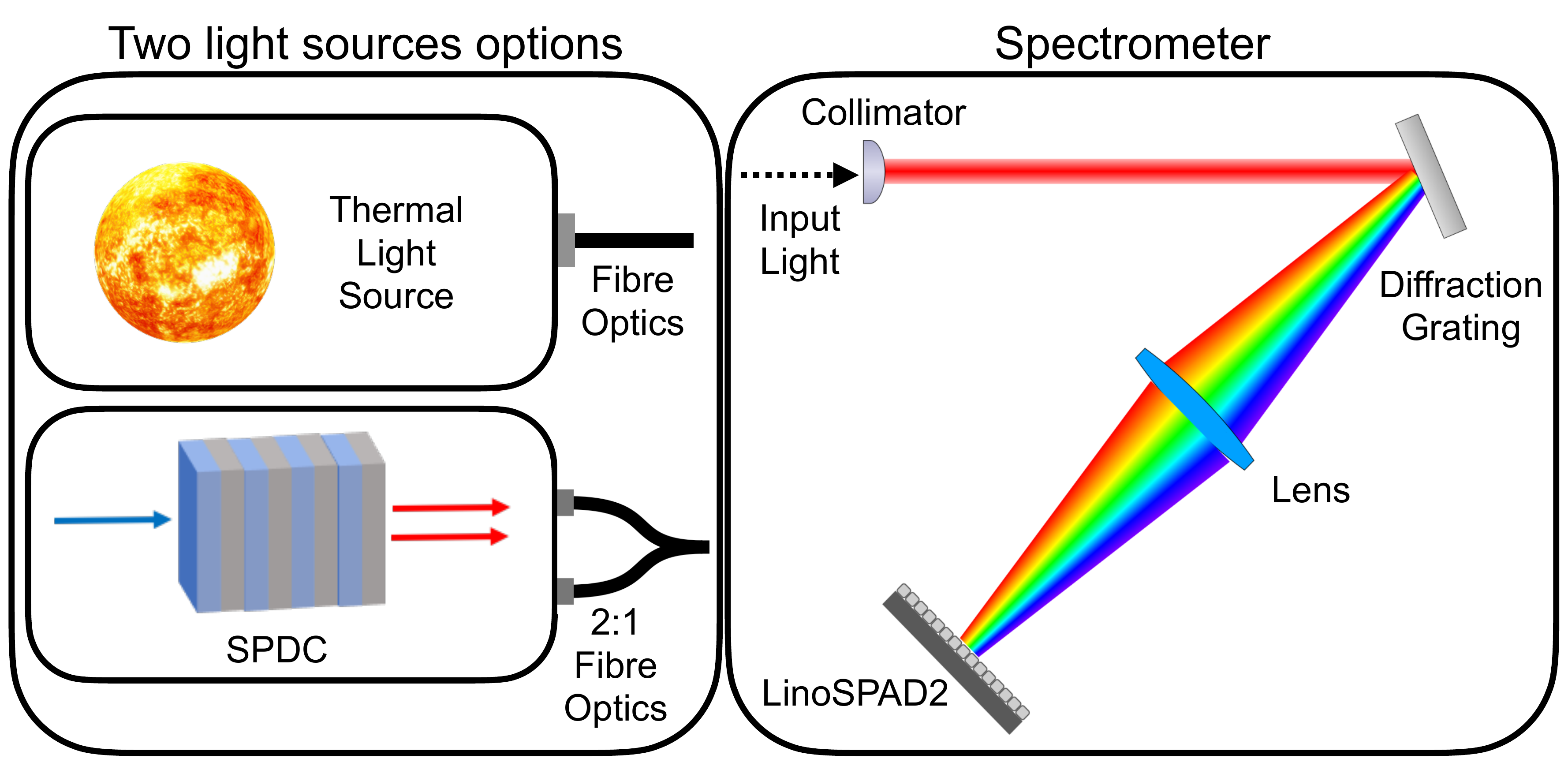}
    \caption{Schematic layout of the spectrometer with two possible light sources used for evaluation of the spectral and temporal resolutions.}
    \label{fig:setup}
\end{figure}

A line of single-photon-sensitive pixels with data-driven readout and excellent timing resolution is ideal for implementing a fast spectrometer for quantum applications. Light enters the setup through a single-mode optical fiber and a collimator, then reflects on a diffraction grating and is focused on the LinoSPAD2 sensor. The entire setup is enclosed in a box to avoid optical background noise. Figure~\ref{fig:setup} depicts a schematic of our experimental setup. We employed two light sources to characterize the spectrometer, a thermal light source and a spontaneous parametric down-conversion (SPDC) single-photon source, as explained in more detail later.


The implemented spectrometer was designed employing a 1200 lines/mm diffraction grating and f = 200~mm lens. Before the grating, the incoming light is collimated with a lens, producing a collimated beam with an approximately 7 mm diameter. The first-order reflections from the grating are focused on the sensor placed at the focal distance from the lens. The resulting wavelength conversion scale is 0.11 nm/pixel.
Alignment of the spectrometer optical setup with respect to the very thin linear LinoSPAD2 array was a difficult task, requiring precision. We are devising improved sensor architectures to alleviate this issue.

\subsection{The LinoSPAD2 sensor}

\begin{figure}[]
    \centering
   \includegraphics[width=1.0\linewidth]{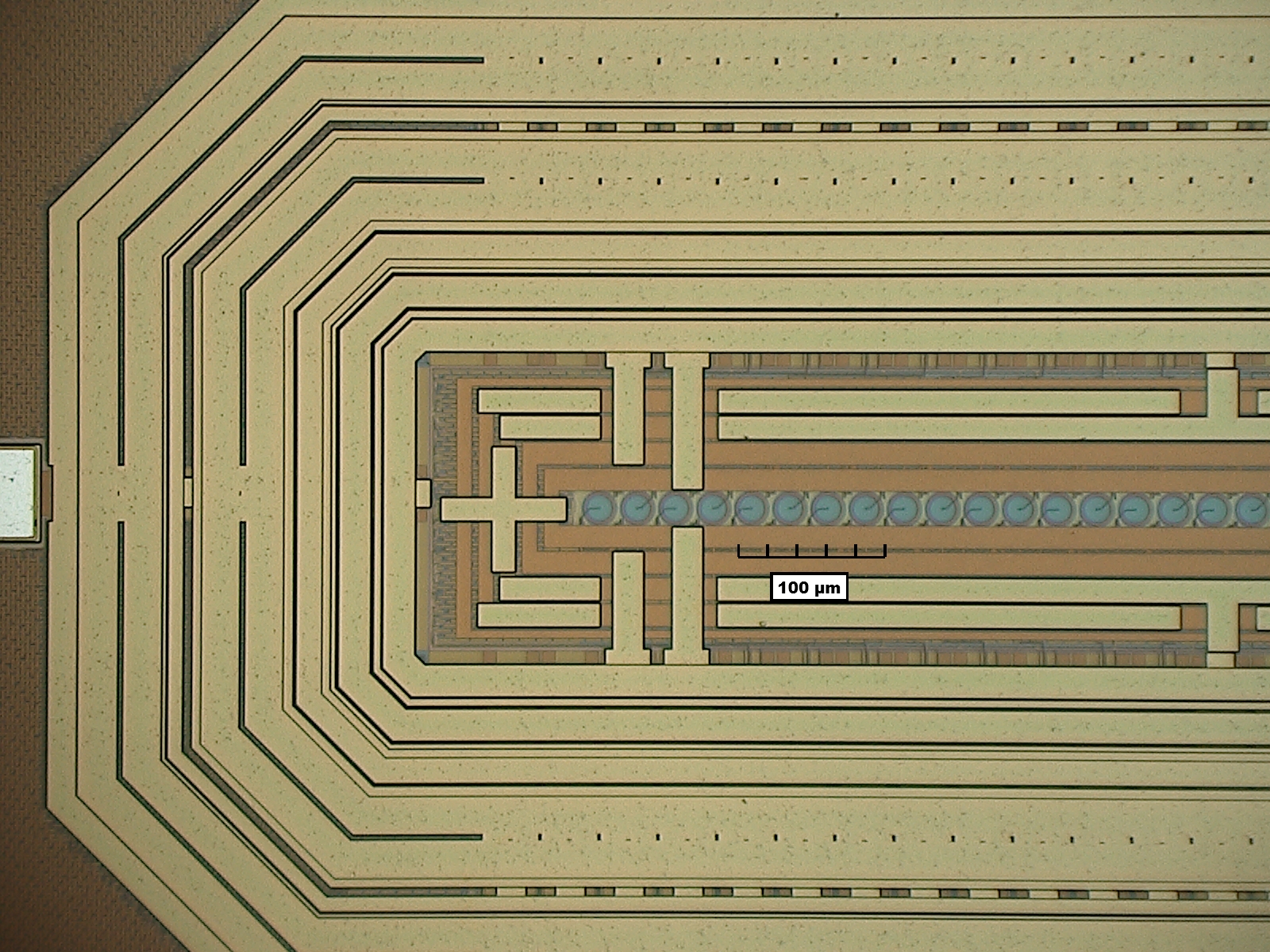}
    \caption{Photograph of the LinoSPAD2 sensor edge. 
    The line of pixels is in the centre, oriented horizontally. 
    Reproduced from Ref.~\cite{milanese2023linospad2}.
    }
    \label{fig:LinoSPAD2}
\end{figure}

In the LinoSPAD2 sensor, each pixel is a single-photon-sensitive photodiode with a $26.2 \text{ µm} \times 26.2$ µm size and 25\% fill factor. The sensor consists of a linear 512x1 pixel array with a median dark count rate (DCR) below 100~Hz per pixel and a peak photon detection probability of approximately 50\%. The use of microlenses achieves an additional improvement in sensitivity by a factor of 2.3 on average, with a final photon detection efficiency (PDE) of about 30\%  at 520~nm \cite{bruschini2023,Bruschini23_microlens}. For the present work, we used only 256 pixels. Each pixel in the array allows for single-photon detection with time-stamping by employing an FPGA. The FPGA implements an array of time-to-digital converters that can be reconfigured in terms of range, and least significant bin (LSB) size or replaced by digital counters for simple photon counting. This reconfigurability permits adjustments in both the readout and processing chains, and in the spatial and temporal granularity, thus allowing one to match specific application requirements. Figure~\ref{fig:LinoSPAD2} presents a photograph of the LinoSPAD2 sensor. 

The full LinoSPAD2 sensor consists of two halves, 256 pixels each, which are independently connected to two Spartan-6 FPGAs. Time stamping is performed by an array of 64 time-to-digital converters (TDC), implemented in the FPGA and shared between 256 channels. The TDCs consist of 35-element delay lines, where each element is made of a 4-bit carry-chain block. The resulting 140-bit time code is sampled and encoded into an 8-bit number, with differential and integral nonlinearities being reduced by calibration of the LSB, re-binning, and by elimination of non-monotonicity. The sampling frequency, derived from an external crystal oscillator, is 400 MHz at nominal conditions. A 20-bit counter operating at the same frequency expands the 8-bit code obtained from the TDC to a 28-bit timestamp with a least-significant bit (LSB) resolution of 17.857 ps. The longest possible acquisition cycle is 4 ms long, for which an internal trigger of 250 Hz is used. The pixel dead time, that is, the time after detection for which a pixel is insensitive to another photon detection, is about 50~ns. 

The FPGA internal block memory can store up to 512 timestamps per pixel per acquisition cycle.
A USB 3.0 interface is used for communication and data exchange between the readout board and the acquisition computer. Given all the above parameters, the sensor and its acquisition system, as was used here, we are able to time stamp a stream of single photons and read them out to a computer with a maximum rate of 8M photons per second.

Possible limitations due to the linearity of the sensor response may arise due to the pixel deadtime of 50~ns. Therefore, the non-linearity may start to influence the measurements above rates of a few MHz per pixel. In our experiments the maximum rates did not exceed 100 kHz per pixel, far below the values where the non-linearity is important \footnote{In this paragraph, we referred to the photon detection rate as in Hz, instead of counts per second. We note that, from a technical perspective, Hz refers to periodic signals, which is not necessarily the case here. In other words, we are referring simply to counts per second.}. 

\subsection{Data analysis and calibration}

The data was analyzed by calculating the time difference between timestamps of different SPAD pixels. Only timestamps from a single acquisition cycle are taken into account. Additionally, due to TDC non-linearities and unequal time offsets for different pixels, the data must be adjusted to account for these effects.

The TDC 140-bit thermometer output code, sampled at 400 MHz, covers a time window of 2.5~ns, with an average bin width of 17.857~ps. However, the actual width of each bin is different due to uneven propagation delays in the FPGA carry chain logic. In addition, due to the different trace lengths between each SPAD and the inputs of the TDCs, there is a static offset in time-stamping that varies from pixel to pixel. Two calibration procedures were therefore required. The first was used to compensate for the non-linearity of each TDC, and the second to compensate for the offset of each SPAD.

The non-linearity of each TDC has been calibrated using a code density test technique, in which the sensor is homogeneously illuminated with a uniform light source. Ideally, this should result in all histogram bins being equally populated. In reality, the bins with larger propagation delays have higher occupancy, and bins with lower propagation delays have lower occupancy. The occupancy of individual bins directly gives the bin time width, assuming that the 140-bit thermometer code covers exactly 2.5 ns. Recalibrating the output data of each TDC compensates for the non-linearity and, thus gives the real-time information of individual timestamps.

The offset of each SPAD was calibrated using a picosecond pulse laser flashing at the entire sensor. With the assumption that all SPADs should register photons at the same time, we compiled and solved a set of linear equations with corresponding time differences for all possible pixel pairs. This allowed us to determine the individual offset delays of each SPAD. The resulting calibration data, non-linearity, and time offsets were then used for the offline data analysis.

\subsection{Spectrometer spectral and temporal resolutions}

\begin{figure}[]
    \centering
    \includegraphics[width=1.0\linewidth]{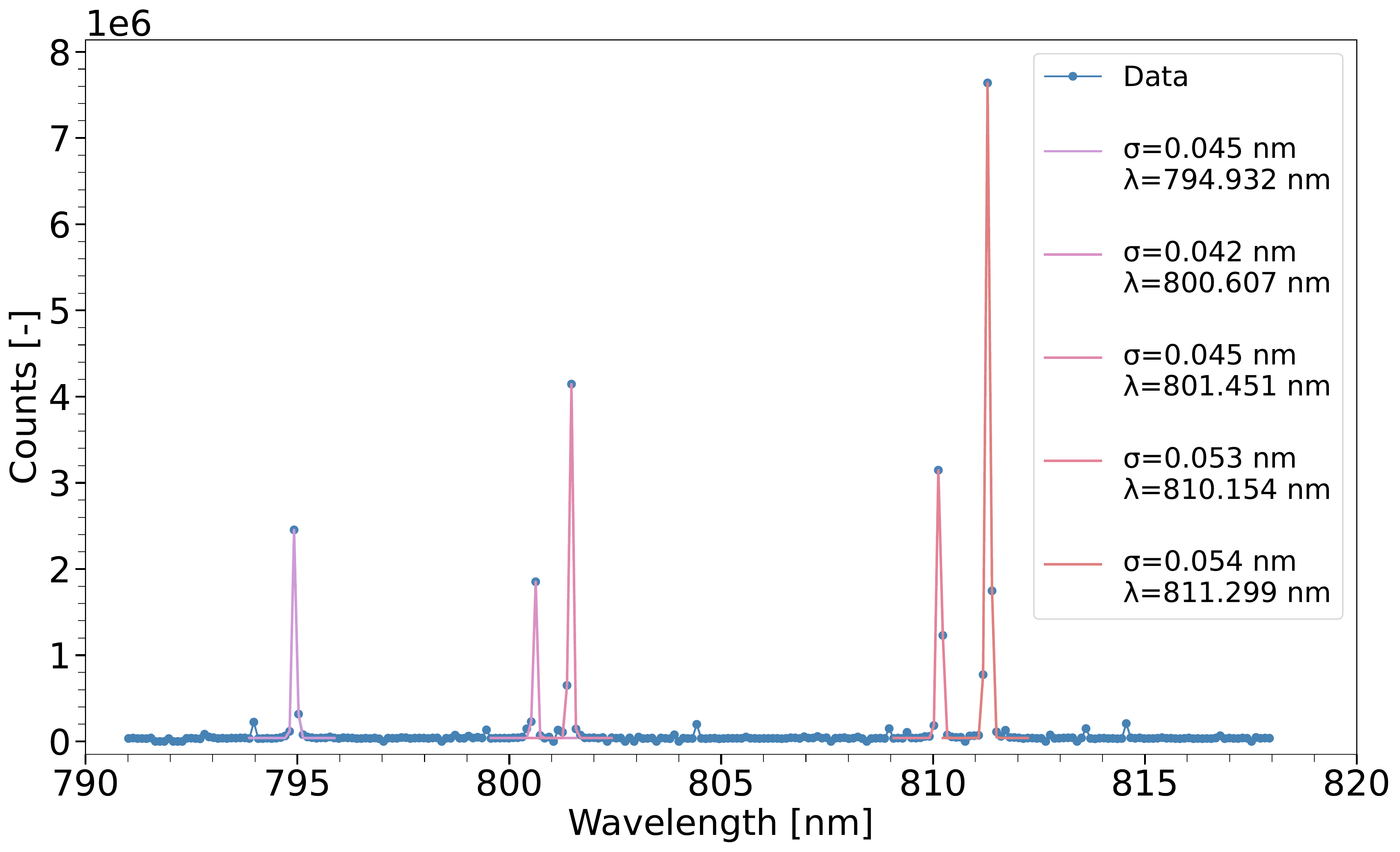}
    \caption{Argon spectrum measured with the fast spectrometer.}
    \label{fig:argon}
\end{figure}

First, we report the spectral resolution. We employed a Newport 6030 Argon calibration lamp at 10~mA DC current as a source of thermal light. The lamp light was coupled into a single-mode fiber. The spectrometer was characterized using the argon emission spectrum~\cite{NIST_ASD}, corrected for transmission through air, which has a large number of narrow lines. A section of the resulting pattern after the diffraction grating is shown in Fig.~\ref{fig:argon}. The peaks correspond to spectral lines of argon, which have well-known wavelengths and, therefore, provide an excellent calibration of the spectrometer. The peaks are fitted with a Gaussian function to determine the spectral resolution (rms). We selected a spectral range of 30 nm centered at 805 nm. Our best result was a spectral resolution of $\sigma=0.042$ nm for the 800.607 nm line, while the spectral resolution varied between 0.042 and 0.054~nm. The total time to acquire the dataset was 400~s.

\begin{figure}[]
    \centering
    \includegraphics[width=1.0\linewidth
    ]{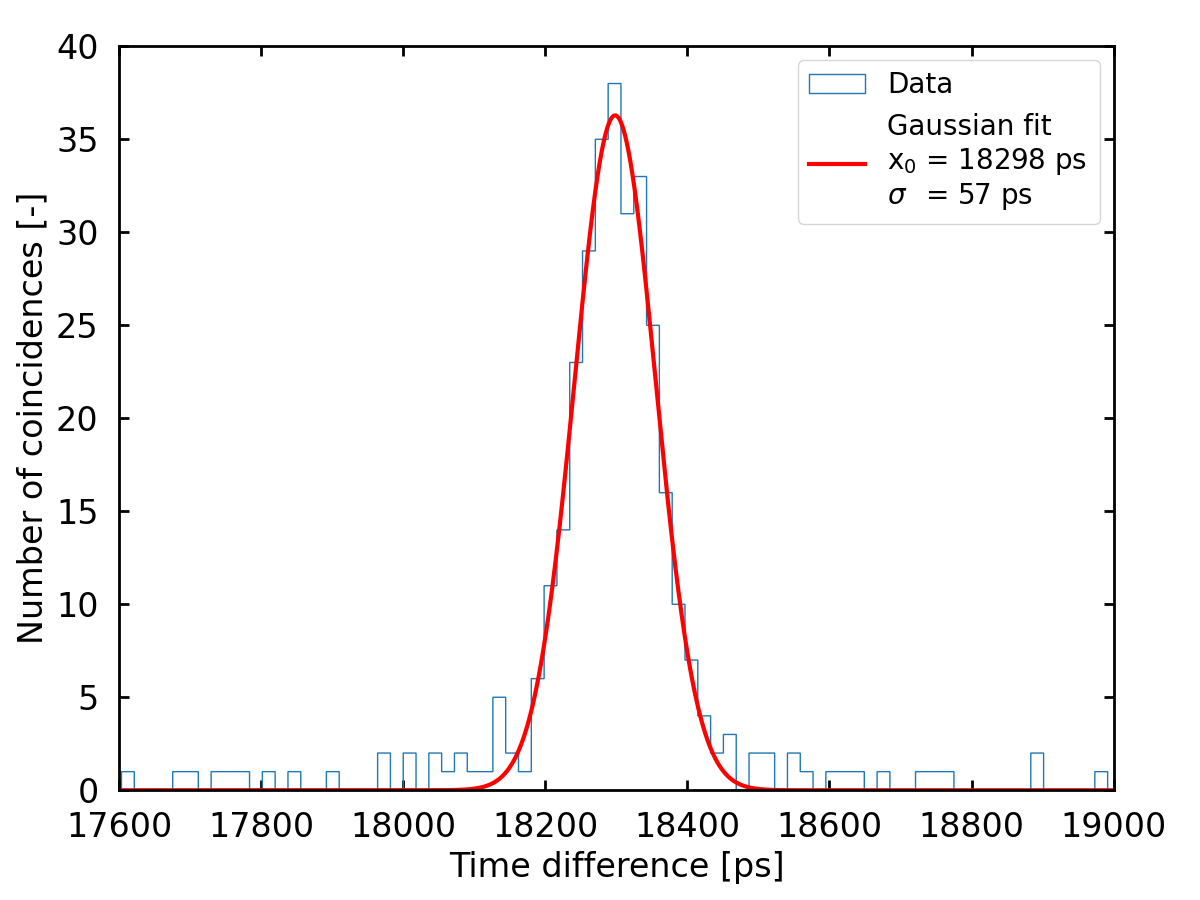}
    \caption{Distribution of time differences for two-photon coincidence detection from an SPDC source.}
    \label{fig:timedifference}
\end{figure}

Second, we now report the temporal resolution. We used a commercial spontaneous parametric down-conversion (SPDC) source of simultaneous photon pairs to characterize the timing resolution of the spectrometer \cite{SPDC_general, SPDC_spectra}. We also used a Thorlabs correlated photon-pair source based on spontaneous parametric down-conversion of 405 nm pump photons in a ppKTP crystal~\cite{SPDC_spectra}. This source provides fiber-coupled photon pairs with the pump photon power range from 0 to 150~mW. Since in the SPDC process the two photons in the photon pair, commonly called signal and idler, are generated at the same instant (simultaneously), it makes SPDC-generated photons ideal for analyzing temporal resolution. The idler and signal photons from the source were fiber-coupled and focused onto two separate locations on the sensor, a different setup than what is shown in Fig.~\ref{fig:setup}. Figure \ref{fig:timedifference} shows the distribution of time differences for two-photon coincidence detection. Fitting with a Gaussian function, we obtain a temporal resolution (rms) of 57 ps for coincident counts. Assuming equal contributions to uncertainty for each of the two photons in the pair, we determine a temporal resolution of $57/\sqrt{2}=40$~ps for a single photon, which is one of the best results ever reported for a single-photon-sensitive spectrometer. The peak is centered at 18.3~ns due to different delays in the two arms of the SPDC source. 

Similar spectrometer designs were used in the past but with different single photon detectors and inferior timing resolution \cite{Zhang2020,Svihra2020, Lubin2021}. 

We note that for the purpose of the resolution evaluation of both the spectral and temporal measurements, we are far away from any quantum limitations. The natural temporal coherence of the argon lines is 150~ps \cite{Nomerotski20} corresponding to a spectral width of picometers, which is much smaller than the measured spectral resolution. At the same time the simultaneity of SPDC photons is in 10's of femtoseconds given the bandwidth of the source \cite{SPDC_spectra}, which is much smaller than the current time resolution.

\subsection{Coincident detection of SPDC photon pairs}
\label{sec:time}

Since the sensor readout is data-driven and as such is completely independent for all pixels, it can register multiple photons simultaneously as long as they are incident on different pixels. To demonstrate this in the spectrometer configuration, the idler and signal photons were combined using a 2:1 fiber splitter and then directed through the spectrometer in the same optical path as the argon thermal light, conceptually depicted in Fig.~\ref{fig:setup}. The SPDC signal and idler have different non-degenerate spectra \cite{SPDC_spectra}. Therefore, the diffraction grating will cause them to reach different pixels in the LinoSPAD2. 

\begin{figure}[]
    \centering
    \includegraphics[width=1.0\linewidth]{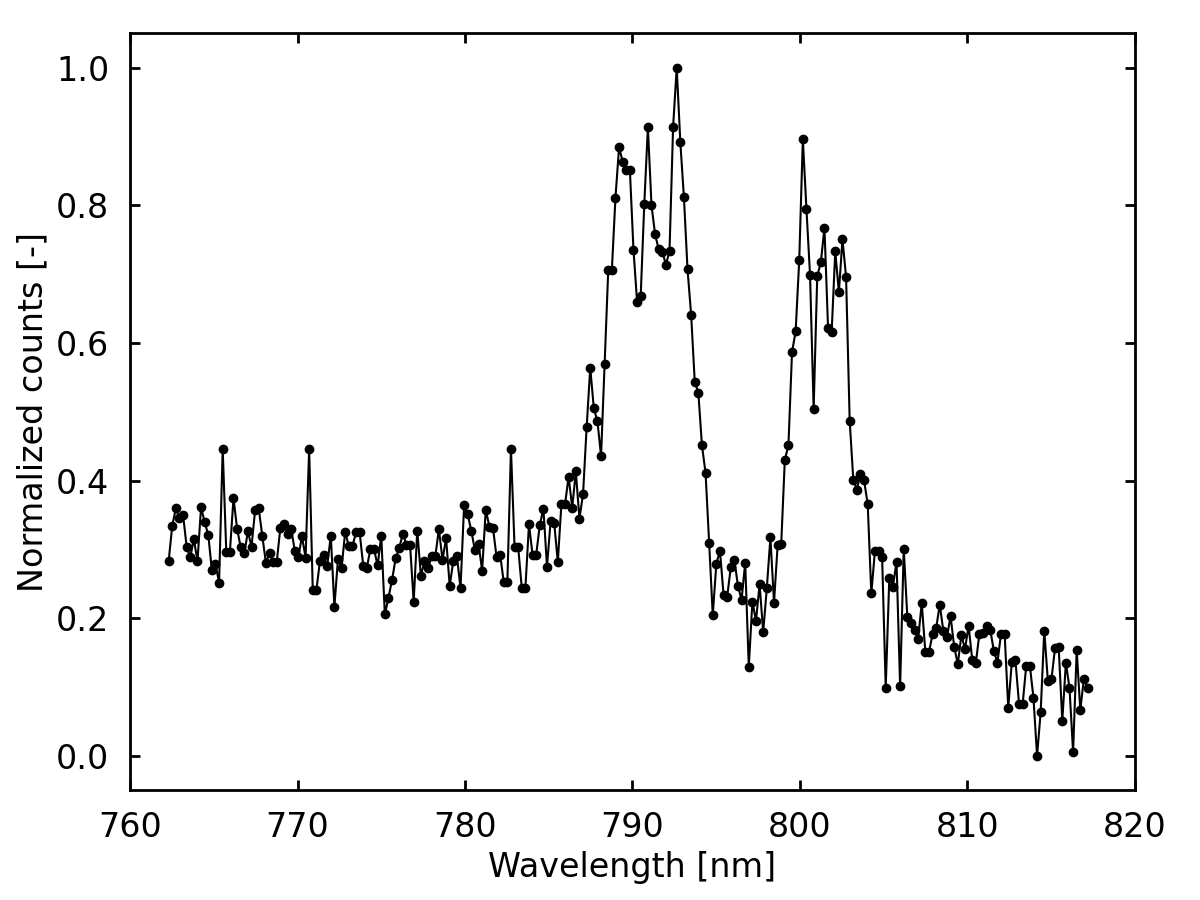}
    \caption{Spectra of signal and idler photons from the SPDC source in the LinoSPAD2 sensor.}
    \label{fig:SPDC50mW}
\end{figure}

To prove that the spectrometer is sensitive to simultaneous pairs of photons, we analyzed their spectrum in our spectrometer. The result is shown in Fig.~\ref{fig:SPDC50mW} for the case of an SPDC pump power of 50 mW. There are two wide peaks in the distribution corresponding to the signal and idler photons. The left and right peaks are the signal and idler photons, respectively, which were determined from previous measurements \cite{SPDC_spectra}.

\begin{figure}[]
    \centering
    \includegraphics[width=1.0\linewidth]{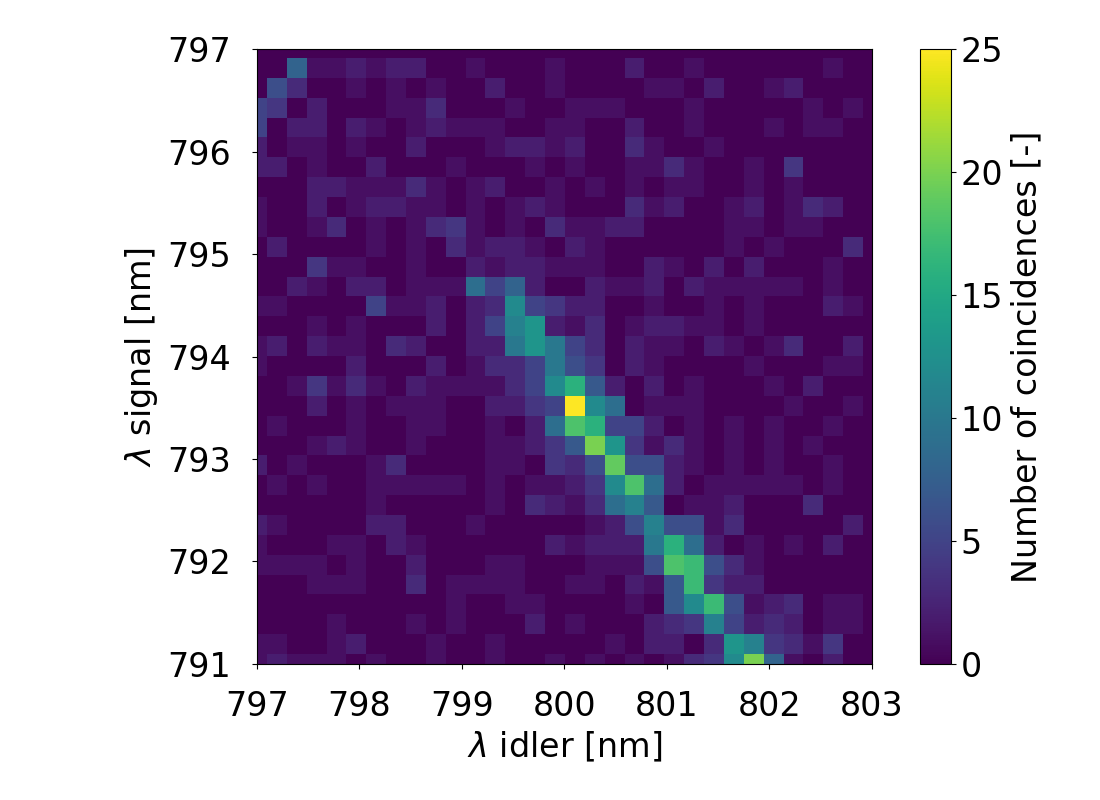}
    \caption{Anti-correlation of wavelengths for coincident signal and idler photons from the SPDC source.}
    \label{fig:anticorreation}
\end{figure}

The SPDC photons present a typical anti-correlation in wavelengths due to the energy conservation, i.e. $\hbar\omega_{\mathrm{pump}}=\hbar\omega_{\mathrm{signal}}+\hbar\omega_{\mathrm{idler}}$. Figure~\ref{fig:anticorreation} shows the wavelength anti-correlation for the signal and idler photons for coincident pairs within a 20~ns time window. The anti-correlation is clear and is a signature of the SPDC process, although more sophisticated SPDC sources may have engineered spectral properties \cite{SPDC_general}.

\subsection{Benchmark to the Heisenberg Uncertainty Principle (HUP)}

\subsubsection{Benchmark}

We benchmark our obtained results to the limit determined by the Heisenberg Uncertainty Principle (HUP), which imposes resolution limits for joint measurement of two conjugate observables~\cite{heisenberg1949physical,sakurai2014modern,hall2013quantum,PRL_1929_Robertson} and is therefore a useful benchmark to determine how close our spectrometer is to the ultimate quantum limit. The Heisenberg Uncertainty Principle does not limit the resolution for one of the observables alone but restricts the product of the uncertainty of two observables. For energy and time, it takes the form of:
\begin{equation}
    \Delta E \Delta t \geq \frac{\hbar}{2},
\end{equation}
where $\hbar$ is Planck's reduced constant, $\Delta E$ is the standard deviation for the measurement of energy, and $\Delta t$ is the standard deviation for the measurement of time. 

The uncertainty in energy can be estimated from the uncertainty in wavelength. Using $\Delta \lambda = 0.042$ nm for the 800.607 nm light and propagating the errors, we obtain $\Delta E = 8.1 \times 10^{-5}$ eV. The temporal uncertainty per photon is 40~ps, as determined before. We find the product of the uncertainties in time and energy to be:
\begin{equation}
     (\Delta E \Delta t)_{experimental} = 3.3\times10^{-15}\ \text{eVs}.
\end{equation}

Therefore, $(\Delta E \Delta t)_{experimental}/(\hbar/2) \approx  10$, which tells us that we are a factor of ten from the Heisenberg Uncertainty Principle limit of $\hbar/2$.

We emphasize that the Heisenberg Uncertainty Principle is applicable to a single particle, and that the spectrometer can simultaneously measure the energy and time for each detected photon, even when multiple photons are detected at the same time, provided the photons reach different pixels. Considering the simplicity and straightforwardness of our experimental setup, this represents a significant result.

An important clarification about the HUP benchmark is needed. We are not measuring photons in such a way as to claim a direct comparison to the HUP. We are using the HUP as a benchmark, a point of reference. We directly measured the spectral resolution using the thermal light, and we directly measured the temporal resolution using SPDC-generated photons. We assume that the same resolutions would apply to other photon generation mechanisms and different wavelengths in a reasonable range. We also assume that the measured spectral and temporal resolutions apply to a single photon. For the spectral resolution, we used a thermal light source, and thus the light is in a classical state and is well represented by a super-poissonian photon-number distribution~\cite{fox2006QO}. Later, we directly measured the temporal resolution using the quantum state of light produced a SPDC photon-pair source, a biphoton quantum state. The HUP applies to a single photon when a simultaneous measurement of energy and time is performed, which is not the scenario here, although our reported device could be used for such measurements. Nevertheless, under the assumptions that our measured spectral and temporal resolutions would be carried to a simultaneous measurement of a single photon, the HUP is the natural benchmark to compare with. Bear in mind that HUP imposes the ultimate limit of what is experimentally achievable for the simultaneous measurement of a single photon's energy and time of detection.

For completeness of the HUP discussion, we must say that the original HUP argument encompasses three ways that the uncertainties of pairs of incompatible observables are associated, as well discussed in Ref.~\cite{ringbauer2017exploring}: "(I) two incompatible observables cannot be arbitrarily well defined on a quantum state; (II) two incompatible observables cannot be jointly measured with arbitrary accuracy; (III) the measurement of one such observable disturbs the subsequent measurement of the other." In our case, we are using the sense in case (II) with the caveats mentioned above. For other discussions of the HUP in terms of error-disturbance uncertainty, see Refs.~\cite{PRA_Ozawa2003, PRA_Hall2004, branciard2013error, PRL_HUP_Brisbane2014,PRL_HUP_Toronto2012, PRL_HUP_Japan2014}.

\subsubsection{Theoretical considerations on the Heisenberg Uncertainty Principle for a biphoton state}

Here we extend the discussion by analysing the biphoton state produced in an SPDC photon source.

A SPDC two-photon state is described in Fock basis as follows:
\begin{multline}
    \ket{\Psi}_{\rm{SPDC}} = \ket{0_{{\rm{s}}}0_{{\rm{i}}}} \\
    + \eta \int \Psi(\omega_{{\rm{s}}},\omega_{{\rm{i}}})\ket{1_{\omega_{{\rm{s}}}}1_{\omega_{{\rm{i}}}}}d\omega_{{\rm{s}}}d\omega_{{\rm{i}}} + O(\eta^{2}), \eta \ll 1,
\end{multline}
where $\eta$ is the probability amplitude associated with the pair generation state $\ket{1_{\omega_{{\rm{s}}}}1_{\omega_{{\rm{i}}}}}$.
We approximate the resulting state as a biphoton state where the exact number of light quanta is equal to two, corresponding to a photon pair, mathematically the state $\ket{1_{\omega_{{\rm{s}}}}1_{\omega_{{\rm{i}}}}}$, with zero variance, assuming that our detectors have $100 \% $ quantum efficiency. Further discussion on the photon number decomposition can be found in Ref.~\cite{SPDC_general}, and an example of its consequence to photonic quantum computing can be found in Ref.~\cite{Till_photon_number}. We expect to see just two detection events that are separated by time $\Delta t$. This time is tailored to the mode structure with different frequencies of a biphoton amplitude $\Psi(\omega_{s},\omega_{i})$. For both signal $s$ and idler modes $i$, we assume that we are using frequency filters with spectral width $\Delta \omega _{f}$ near half of the pump frequency $\omega _{p}$. Thus, the biphoton amplitude has the following form:

\begin{equation}
\label{biphotonampl}
\Psi(\omega_{s},\omega_{i}) \propto e^{-\frac{(\omega_{\rm{p}} - \omega_{\rm{i}} - \omega_{\rm{s}})^{2}}{2\Delta \omega^{2}_{\rm{p}}}}e^{-\frac{(\omega_{\rm{i}}- \omega_{p}/2)^{2}}{2\Delta \omega^{2}_{\rm{f}}}}e^{-\frac{(\omega_{\rm{s}}- \omega_{p}/2)^{2}}{2\Delta \omega^{2}_{\rm{f}}}}, 
\end{equation}

 The model of the biphoton amplitude can be simplified further by employing Gaussian functions with effective width:

\begin{equation}
\label{biphotonampl}
\Psi(\omega_{s},\omega_{i}) = \frac{1}{\sqrt{\pi \Delta \omega_{pe} \Delta \omega_{ce}}}e^{-\frac{(\omega_{\rm{pe}} - \omega_{\rm{i}} - \omega_{\rm{s}})^{2}}{4\Delta \omega^{2}_{\rm{pe}}}}e^{-\frac{(\omega_{\rm{i}} - \omega_{\rm{s}})^{2}}{4\Delta \omega^{2}_{\rm{ce}}}}, 
\end{equation}

where $\Delta\omega_{pe}$ and  $\Delta \omega_{ce}$ are effective widths. The  $\Delta\omega_{pe}$ is defined as follows: $\Delta\omega_{pe} = \sqrt{\frac{\Delta \omega _{p}^{2}}{2} + \Delta \omega _{f}^{2}}$, $\Delta \omega_{ce} \equiv \Delta \omega _{f}$.
For this state, one can estimate a variance of energy in a state $\ket{\Psi}_{\rm{SPDC}}$:

\begin{equation}\label{en_var}
    {[\sigma (E)}]^{2} = \bra{\Psi} \hat{H}^{2}\ket{\Psi}_{\rm{SPDC}} - (\hbar \omega_{pe})^{2}  = \hbar^{2}(\Delta\omega_{pe}^{2}),
\end{equation}

On the other hand, the variance of $\Delta t$ between detections of two photons can be evaluated via the width of second order correlation function in the time domain, expressed in the following standard form: 
\begin{multline}
        \Gamma(t,t+\Delta t) \approx || \hat{E}_{s}(t)\hat{E}_{{\rm{i}}}(t + \Delta t)\ket{\Psi}_{\rm{SPDC}}||^{2} = \\
        |\tilde{\Psi}(t,\Delta t)|^{2} \equiv |\frac{1}{2\pi}\int \Psi(\omega_{{\rm{s}}},\omega_{{\rm{i}}})  e^{-i\omega_{{\rm{i}}} t} e^{- i \omega_{{\rm{s}}} (t + \Delta t)}d\omega_{{\rm{i}}}d\omega_{{\rm{s}}}|^{2},
        \label{corr}
\end{multline}
where $\hat{E}_{\rm{i(s)}}(t) = \frac{1}{\sqrt{2\pi}}\int \hat{a}_{\rm{i(s)}}e^{-i\omega_{i}t}$ is a positive frequency electrical field operator. 
From the above, one can estimate the time difference \mbox{variance} for the biphoton coincidence detection:
\begin{equation}
     [\sigma (\Delta t)]^{2} = \int (\Delta t)^{2} |\tilde{\Psi}(t,\Delta t)|^{2} d\Delta t dt =  \left(\frac{1}{\Delta \omega_{ce}}\right)^{2}
\end{equation}
Thus, the time resolution  $\sigma (\Delta t)$ of simultaneous \mbox{detection} of biphotons with energy uncertainty $\sigma (E)$ are connected via the HUP-like limit as:

\begin{equation}\label{UP}
    \sigma (E)\sigma (\Delta t)  = \hbar\left(\frac{\Delta \omega _{pe}}{\Delta \omega _{ce}}\right) = \hbar\left(\sqrt{1 + \frac{1}{2}\frac{\Delta \omega _{p}^{2}}{\Delta \omega_{f}^{2}}}\right)
\end{equation}
 
For the considered biphoton state, the similarity with HUP for non-relativistic particles is straightforward. As we described above, it has a clear physical meaning as it relates the time resolution of coincidence detection of two-photon events $\sigma (\Delta t)$ to the energy uncertainty $\sigma (E)$.

This approach is applicable to arbitrary quantum states of light with the case of two-mode quantum states explicitly worked out above in detail. We also emphasize that the mode structure plays an important role here, since we describe our state of light as a superposition of monochromatic plane waves. In our case, it corresponds to a multidimensional Fourier transformation (Eq.~\ref{corr}) as long as operate with biphoton states. Thus, it is natural that the width of the biphoton amplitude in the temporal domain and the width of its Fourier image will be connected via the Fourier limited uncertainty principle (Eq.~\ref{UP}). This is another way to consider the results.

\section{Conclusion}

We have demonstrated a single-photon-sensitive spectrometer with 0.04 nm spectral and 40 ps temporal resolution, based on the LinoSPAD2 sensor. The spectrometer resolution makes it a unique instrument capable of characterizing properties of single photons near the Heisenberg Uncertainty Principle limit, only ten times above it.

We note that the spectral resolution could be improved by employing echelle gratings in which a standard first-order grating is placed perpendicularly to a second grating. High resolution is provided by the high-order diffraction, and the first-order grating vertically separates the overlapping modes. The output of an echelle spectrometer is a series of parallel stripes; a two-dimensional array of single-photon detectors should be used to give high spectral resolution across a large range. The typical resolution for this type of spectrometer is around 0.01 nm, a possible improvement by a factor of four over our results.

The SPAD technology is also quickly progressing, improving the parameters relevant to the measurements described here: photon detection efficiency, timing resolution, and array size. In particular, a 7.5~ps (FWHM) timing resolution was demonstrated recently \cite{Gramuglia2022, Gramuglia2022_1}, more than a factor of 10 superior than measured in this work, if we convert FWHM to rms resolution used here. Two-dimensional 1~Mpixel SPAD arrays were produced and tested  \cite{Optica2020_EPFL}, though not with the data-driven readout as reported here. Fully reconfigurable two-dimensional SPAD-based cameras would ultimately be built using 3D-stacking technologies, whereby SPADs would be employed as photon-to-digital conversion elements, stacked onto CMOS-based reconfigurable circuitry in an FPGA-like bottom tier \cite{bruschini2023}.

Examples of other spectrometers using a 1D pixel array can be found in references~\cite{Grabarnik2008,
Jennewein_spectr2014SR,Jennewein_spectr2014JAP, Lubin2021}. Although these works are relevant in their own right, the present work achieves a considerably better performance.

Our work can help characterize single-photons sources~\cite{SPDC_general, senellart2017high, JMP_2018_Joint-spectral-characterization}, in particular their spectral properties and their implications for quantum interference~\cite{HOM_effect,frequencyHOM,PRA_Jordan2022}. Other areas that could also benefit are time-resolved spectroscopy~\cite{Lubin2021}, quantum metrology~\cite{advances_qmetrology}, time-domain quantum optical modes~\cite{raymer2020temporal}, and quantum spectroscopy~\cite{roadmap_qspect}. The technology we demonstrate here has the potential to unlock many applications in both classical and quantum optics.

\begin{acknowledgments}

This work was supported by the U.S. Department of Energy QuantISED award, the Brookhaven National Laboratory LDRD grant 22-22, the Ministry of Education, Youth and Sports of the Czech Republic Grant No. LM2023034, as well as from European Regional Development Fund-Project ``Center of Advanced Applied Science" No. CZ.02.1.01/0.0/0.0/16-019/0000778. This work was also supported by the EPFL internal IMAGING project ``High-speed multimodal super-resolution microscopy with SPAD arrays" and the DOE/LLNL project ``The 3DQ Microscope". We are grateful to Duncan England, Yingwen Zhang, Dmitri Kharzeev, Rene Reimann, Konstantin Katamadze, and Paul Stankus for useful discussions.

\end{acknowledgments}

\bibliography{bib}
\bibliographystyle{ieeetr}

\end{document}